\documentclass[aps,pra,twocolumn,superscriptaddress,floatfix]{revtex4}
\usepackage{amsmath,amssymb,graphicx,graphics,color}
\usepackage{dcolumn}
\usepackage{bm}
\usepackage{psfrag}
\usepackage{epstopdf}

\begin{document}

\title{Magnetic field induced band insulator to Mott insulator transformations in 4-component alkali fermions at half-filling}
\author{J. Jaramillo}
\affiliation{Institut f\"ur Theoretische Physik, Leibniz Universit\"at Hannover, 30167~Hannover, Germany}
\author{S. Greschner}
\affiliation{Institut f\"ur Theoretische Physik, Leibniz Universit\"at Hannover, 30167~Hannover, Germany}
 \author {T. Vekua}
\affiliation{Institut f\"ur Theoretische Physik, Leibniz Universit\"at Hannover, 30167~Hannover, Germany}

\begin{abstract}
Under the influence of an external magnetic field and spin-changing collisions, the band insulator (BI) state of one-dimensional (1D) $s$-wave repulsively interacting 4-component fermions at half-filling transforms into Mott insulator (MI) states with spontaneously broken translational symmetry: a dimerized state for shallow lattices and a N{\'e}el state for deep lattices via an intermediate topological state. Since a BI has vanishing entropy per particle, these MI phases could be particularly inviting for experimental realization under the similar conditions as those for $^{40}$K atoms \cite{Sengstock}, provided the magnetic field is changed adiabatically. 
\end{abstract}

\maketitle

\date{\today}




Ultra-cold spinor Fermi gases in optical lattices are a potentially extraordinary tool for the modelling of strongly-correlated condensed matter systems under well-controlled conditions \cite{Lewenstein2006,Bloch2008}. The metal-to-MI transition was experimentally observed with ultra-cold 2-component Fermi gases \cite{Jordens2008,Schneider2008}. The MI phase should acquire a magnetic N\'eel (antiferromagnetic) ordering at sufficiently low temperatures, due to the super-exchange interaction, which at the moment has not yet been resolved experimentally \cite{Jordens2010}, the main obstacle being the absence of efficient cooling methods in strongly correlated regimes in the presence of an optical lattice \cite{DeMarco}.  With increasing number of components new interesting physics emerges as multi-component lattice Fermi gases offer possibility of hosting exotic ground state, like 2D spin liquids \cite{Hermele} or 1D topological states \cite{Nonne,Totsuka}.

Dynamical properties of multi-component ultra-cold fermions have attracted recent experimental and theoretical attention as well. Spinor dynamics were studied in the case of $s$-wave interacting 4-component fermions both experimentally \cite{Sengstock} and theoretically within mean-field approximation \cite{Dong}. In recent experiment \cite{Sengstock} initially a state of 2 atoms per site was stabilized by the quadratic Zeeman effect; the starting state being a BI of the 2-component Fermi Hubbard model. Although the $^{40}$K atoms were loaded in a 3D lattice, hopping was allowed along one direction, and since $s$-wave interactions are only felt on-site the system studied experimentally was collection of 1D decoupled lattices. After quenching the magnetic field, multiflavour spin dynamics and strong damping due to many-body effects were observed. 

In this letter we analyse the ground state phases of 4-component alkali fermions at half-filling using non- perturbative analytical and numerical tools, particularly tailored for studying 1D systems. In the parameter space of magnetic field and lattice depth we identify various MI phases: those, that break translational symmetry and are characterised by local order and a topological Haldane phase which does not break any microscopic symmetry and is characterised by non-local string order. Even though these different MI phases are manifested below the ultra-low temperatures the fact that one can start from the BI state which has vanishing entropy per particle, and by adiabatically changing the magnetic field enter into these non-trivial MI phases, makes the system of half-filled 4-component alkali fermions a particularly attractive candidate to resolve ground state spin order in experiments on ultra-cold lattice gases. Note that the spin-changing collision processes, crucial for BI to MI transformations, are maximally pronounced at half-filling corresponding to occupancy of 2 particles per site.


\begin{figure}[t]
\vspace*{0.3cm}
\includegraphics[width=7.8cm]{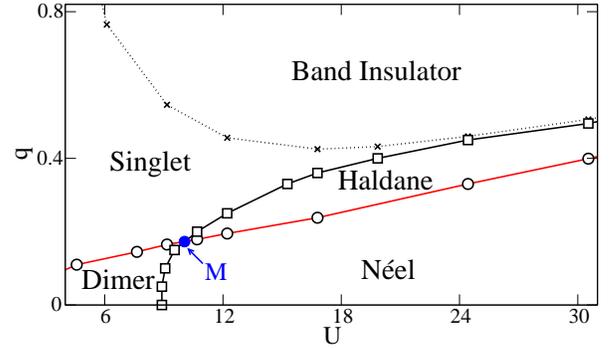}
\caption{(Color online) Numerical ground state phase diagram for a 4-component mixture of $^{40}K$ atoms at half-filling  with average interaction energy $U\sim \frac{4\pi \hbar^2}{m}(a^K_0+a^K_2+a^K_4+a^K_6+a^K_8)/5$ ($a^K_0,...a^K_8$ being the $s$-wave scattering lengths of $^{40}K$) and quadratic Zeeman coupling $q$ (both in units of hopping $t$ set by lattice depth). Dotted line indicates sharp cross-over and continuous lines are quantum phase transition lines, crossing at multicritical point M. For nature of different phases and phase transitions consult the text.}
\vspace*{-0.3cm}
\label{Fig:Main}
\end{figure}

\begin{figure}[t]
\vspace*{0.3cm}
\includegraphics[width=4.8cm]{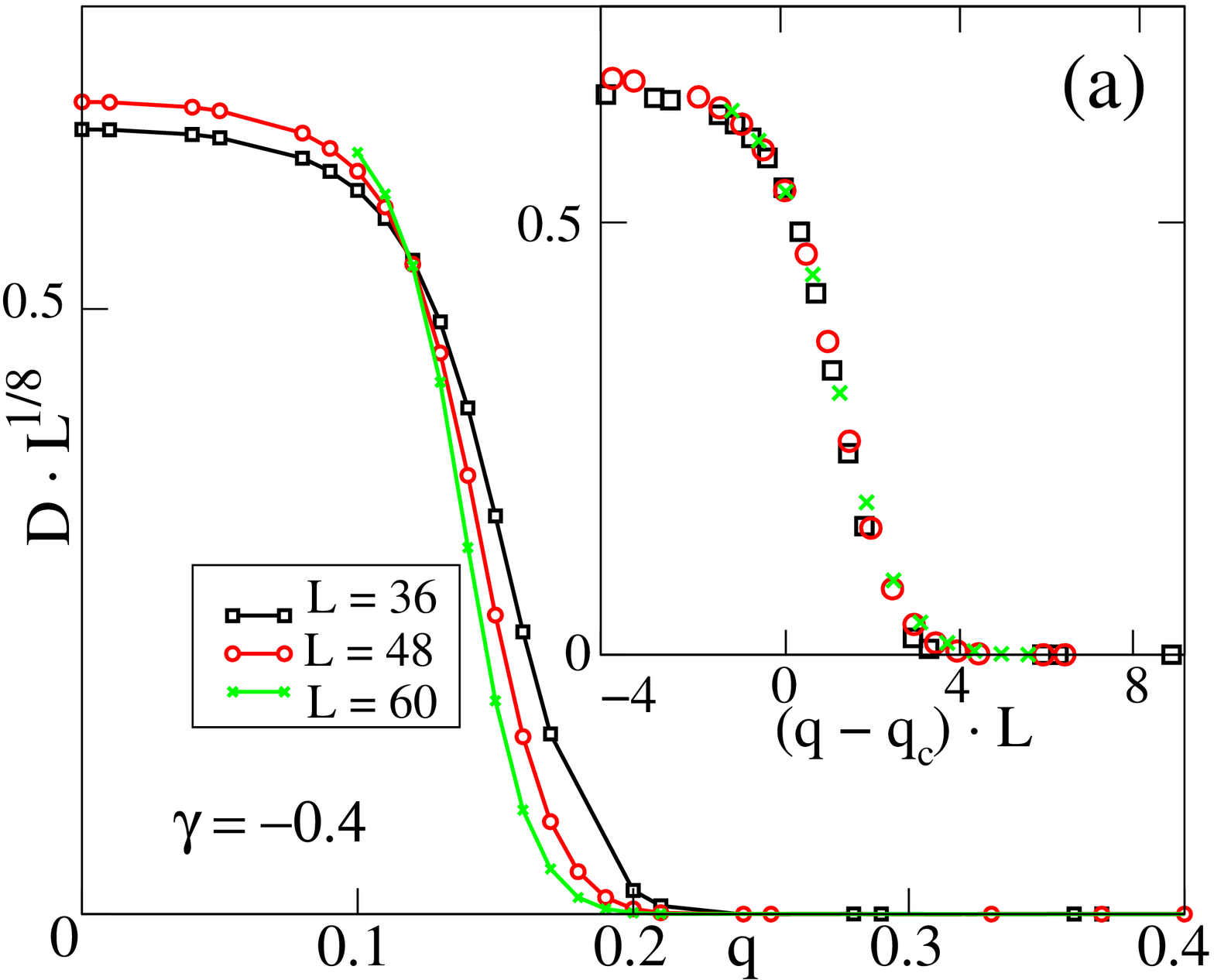}\includegraphics[width=4.0cm]{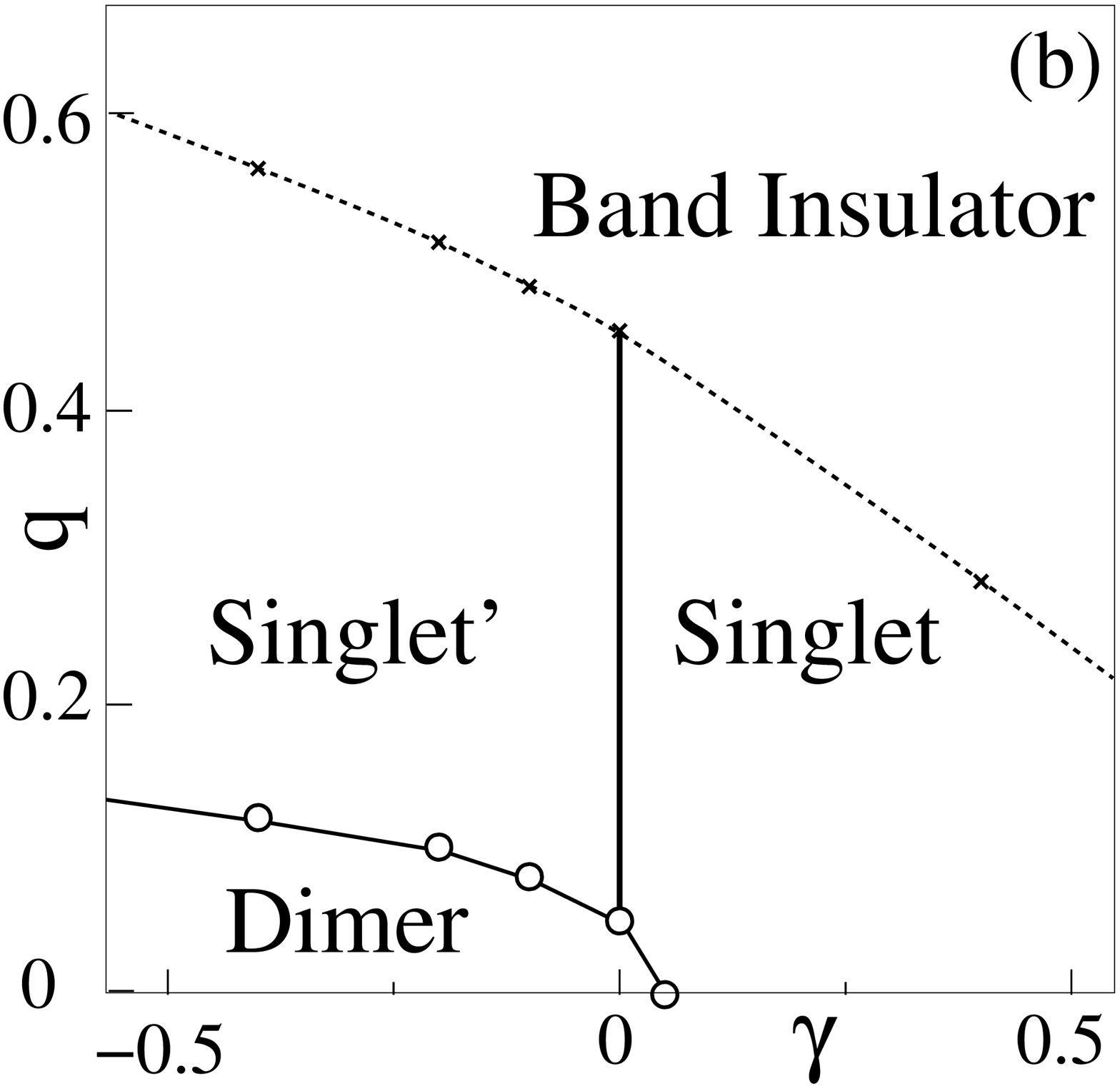}
\caption{(Color online) (a) Scaling of the dimer order parameter for different system sizes. The inset shows the collapse of all finite-system-size data to a single curve, confirming Ising criticality. (b)Ground state phase diagram for $f=3/2$ alkali fermions at half filling, as function of $q$ (in units of $t$) and $\gamma={(g_2-g_0)}/{2t}$ for the case of ${(g_2+g_0)}/{2}=10t$. }
\vspace*{-0.3cm}
\label{Fig:Ising}
\end{figure}

Our main result, the ground state phase diagram of 4 components of $^{40}$K atoms, is presented in Fig.\ref{Fig:Main}. The most relevant finding is that for deep lattices a topological Haldane state developes next to the BI state. This can be important for experiments because 
one can arrive at a topological state starting from the BI of the two component mixture and activate spin-changing collisions by adiabatically changing the magnetic field. This phase diagram is not particular of $^{40}$K atoms, it captures generic phases of repulsive 4-component alkali fermions at half-filling, provided these components do not form complete hyperfine multiplet. For the case of hyperfine spin $f=3/2$, due to large internal symmetry which holds even for nonzero quadratic Zeeman coupling \cite{Vekua}, the situation simplifies and only dimer and singlet/BI phases are realized as depicted in Fig.\ref{Fig:Ising}(b). 




 {\it f=3/2 multiplet-} We start from the case of a hyperfine spin $f=3/2$ \cite{footnote1}. 
The lattice Hamiltonian that describes $s$-wave interacting fermions in the presence of quadratic Zeeman coupling $q$ is a sum of parts quadratic and quartic in fermions, 
\begin{eqnarray}
\label{model}
&&H=\sum_{j=1}^L H_{0,j}+\sum_{j=1}^LH_{int,j},\\
&&\,\, H_{0,j}= \! \!\!\sum_{\alpha=-{3}/{2}}^{{3}/{2}} [-t\, (\psi_{\alpha,j}^\dag\psi_{\alpha,j+1}+ h.c.) +q\alpha^2\,    n_{\alpha,j}], \nonumber\\
&&\,\, H_{int,j}= \sum_{F,|m_F|\le F}\!\!\!  g_F\ P_{F, m_F}^\dag(j) P_{F,m_F}(j).\nonumber
\end{eqnarray}

The operator $\psi_{\alpha}$ annihilates an atom in the hyperfine spin state $|f=3/2,m_f=\alpha\rangle$ and $n_{\alpha,j}=\psi_{\alpha,j}^\dag \psi_{\alpha,j}$.
Interaction coefficients $g_ {F}$ are proportional to scattering strengths $\sim  4\pi \hbar^2a_F/m$ \cite{Jaksch}, where $m$ is the atomic mass, $a_F$ are the $s$-wave scattering lengths and $ P_{F, m_F}^\dag(j)=\sum_{\alpha,\beta}\langle F,m_F|\alpha\beta\rangle\psi_{\alpha,j}^\dag\psi_{\beta,j}^\dag$ create on-site pairs with total spin $F$ and projection $m_F$.

Due to Pauli principle only channels with $F=0,\ 2$ are allowed in the low energy scattering.
Interactions are assumed to be repulsive and we concentrate on the case of two fermions per lattice site on average and zero net magnetization $N_{\alpha}=N_{-\alpha}$, where $N_{\alpha}=\sum_j \langle n_{\alpha,j} \rangle$.

Usually, in alkali metals, differences between the $s$-wave scattering lengths are a few percent of the average scattering length, hence deviations from $SU(4)$ symmetry ($a_0=a_2$ case) are small. At half-filling a unique gapped phase, with spontaneously doubled lattice constant and dimer (spin-Peierls) order, is realized with an order parameter defined as $D= |\sum_{\alpha,j}(-1)^j\langle \psi_{\alpha,j}^\dag\psi_{\alpha,j+1} \rangle |/(L-1)> 0$, in the vicinity of $SU(4)$ point in the absence of magnetic field  \cite{Marston,Solyom2}. We show that for $a_0 \neq a_2$ the magnetic field above a critical value restores translational symmetry via quantum phase transition belonging to the second order Ising universality class and the system enters a site-singlet state that is adiabatically connected to the BI of the two-component Fermi Hubbard model. For the case $a_0=a_2$, relevant for alkaline-earth atoms \cite{Gorshkov}, the dimer state is separated from BI with gapless Luttinger liquid state.

For the 2-component repulsive Hubbard model at half filling the charge sector is gapped and low energy effective theory is described by spin-1/2 Heisenberg chain. In the case of 4-component fermions, at half filling, similarly the charge sector is gapped and instead of one spin sector there are 3 flavour sectors  described by the dual bosonic fields [$\phi_v$, $\theta_v$], [$\phi_{t1}$, $\theta_{t1}$], and [$\phi_{t2}$, $\theta_{t2}$], with $[\theta(x),\partial_y \phi]=i\delta(x-y)$, all of which are generically gapped \cite{Wu06}. The quadratic Zeeman effect couples only to the chiral sector $\phi_v$ . Chirality is defined as $\tau= \frac{1}{2L}\sum_j \langle  [n_{\frac{1}{2},j}+n_{-\frac{1}{2},j} - n_{\frac{3}{2},j}- n_{-\frac{3}{2},j} ] \rangle \sim \langle \partial \phi_v\rangle$, and it measures imbalance between $\pm \frac{1}{2}$ and $\pm \frac{3}{2}$ components. 
At half-filling with a relevant $4k_F$ Umklapp, the important low energy sector is governed by the following Hamiltonian density,
\begin{eqnarray}
\label{effective}
\mathcal{H}_v&=&\frac{v_{v}}{2} \left\{K_{v}(\partial_x\theta_{v})^2+\frac{1}{K_{v}}(\partial_x\phi_{v})^2\right\} \nonumber\\
&+&g\cos\sqrt{4\pi}\phi_v +g_{\gamma}\cos\sqrt{4\pi}\theta_v-\frac{2q}{\sqrt{\pi}}\ \partial_x\phi_v,
\end{eqnarray}


where, $v_v=\sqrt{v_F^2+(g_0-3g_2)v_F/2\pi}$, $K_v=\sqrt{2\pi v_F/(2\pi v_F+g_0-3g_2)}$, $v_F$ being the Fermi velocity of free fermions. Coupling constants $g\sim g_0+ g_2$ and $g_{\gamma}\sim g_2-g_0$, include proportionality coefficients involving averages of massive charge and $\phi_{t_{1,2}}$ fields \cite{Wu06}.

\begin{figure}[ht]
\vspace*{0.3cm}
\includegraphics[width=6.0cm]{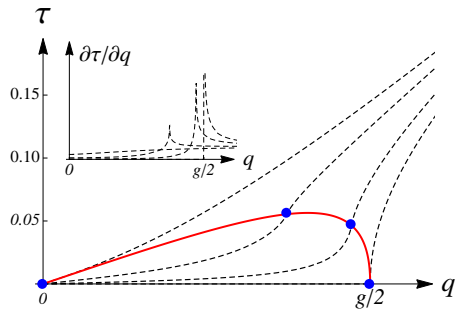}
\caption{(Color online) Chirality versus quadratic Zeeman coupling (dashed lines) for different values of $g_{\gamma}$  ($g_{\gamma}=0$ for the right most curve and $g_{\gamma}=g$ for the left most curve). Intersections of the continuous line with the dashed lines give $q_c$ and critical value of chirality. Inset shows chirality susceptibility and for right most curve the divergence corresponds to commensurate-incommensurate phase transition. }
\vspace*{-0.3cm}
\label{fig:1}
\end{figure}

The effective model in Eq.(\ref{effective}) is diagonalized by refermionization procedure \cite{Gogolin} and Bogoliubov transformation. The low energy excitation spectrum allows us to locate quantum phase transition and determine its nature. In general for $g_{\gamma}\neq 0$ the excitation spectrum is gapped, except at a critical value of the $q=q_{c}$. The low energy spectrum is given by two fermionic branches,\\
$
\omega_{\pm}(k)=\!\!\sqrt{g^2\!+g_{\gamma}^2\!+4q^2+k^2 \pm2\sqrt{g^2g_{\gamma}^2+4q^2(g^2+k^2)}}.
$
Only one  branch becomes gapless for $q_{c}=\sqrt{g^2-g_{\gamma}^2}/2$,  $\omega_{-}(k)=v_-|k|+O(k^2)$. The velocity of the linear mode, at the criticality is $v_-={g_{\gamma}}/{g}$ and the transition belongs to the 2D Ising universality class \cite{YW03}. For $g_{\gamma}=0$ (with additional $U(1)$ symmetry corresponding to conservation of chirality), at $q=q_{c}=g/2$, the linear velocity vanishes and the Ising transition transforms into commensurate-incommensurate one \cite{Gogolin} with characteristic quadratic (non-relativistic) low energy dispersion $\omega_{-}(k)\simeq k^2/2g$.
Dimer phase that is realized for $q<q_{c}$ transforms into singlet phases for $q>q_{c}$. Two singlet phases, corresponding to $g_{\gamma}>0$ and $g_{\gamma}<0$ cases respectively, are separated by a Gaussian phase transition between each other at smaller values of $q>q_c$, however both of them evolve adiabatically into BI state with increasing $q$.
The singlet phases can be approximated as $|S\rangle\sim \prod_j|S\rangle_j$, where $|S\rangle_j=| \frac{1}{2} \rangle_j \otimes| -\frac{1}{2} \rangle_j   - \zeta_{\pm} | \frac{3}{2} \rangle_j \otimes| -\frac{3}{2} \rangle_j $, with $\zeta_{\pm}\to \pm 1$ for $q\to 0$ and $|\zeta_{\pm}| \to 0$ after the crossover into BI. The dependence of chirality and chirality susceptibility obtained from the effective model Eq.(\ref{effective}) at the dimer to singlet phase transitions for different values of $g_{\gamma}$ is depicted in Fig.\ref{fig:1}.

To check the phase diagram at strong couplings (relevant for experiments) we use density matrix renormalisation group (DMRG) calculations for open boundary conditions \cite{White}. We keep in average up to 800 states and $L=60$ sites, comparable to the number of occupied lattice sites per chain \cite{Sengstock}. We monitor directly dimer order parameter $D$. The scaling of dimer order and a collapse of the data for different system sizes on a single curve depicted in Fig.(\ref{Fig:Ising}) (a) confirms Ising character of the dimer to singlet phase transitions $D\sim |q-q_{c}|^{1/8}$. The numerical ground state phase diagram for $f=3/2$ fermions at half-filling is depicted in Fig.\ref{Fig:Ising}(b). Fidelity susceptibility (not shown) has a pronounced peak, both at dimer to singlet phase transition as well as at the singlet to BI crossover, however for the latter case the height does not scale with system size, indicating absence of phase transition (but rather the presence of a sharp crossover) between singlet and BI states. In the BI state chirality (quasi) saturates $\tau\simeq 1$.

The Ising phase transition between dimer and singlet phases, induced by an external magnetic field, generalizes the one that was predicted for $q=0$ using bosonization analysis \cite{Wu06} when changing the scattering lengths $a_0$ or $a_2$ (in case of attractively interacting $f=3/2$ fermions similar phase transition happens between singlet and quarteting phases \cite{Wu03,Lecheminant1,Lecheminant2}). Our work suggests that this phase transition can be induced by external magnetic field via quadratic Zeeman coupling, without the need to change scattering lengths, thereby presenting certain advantages for experiments.

{\it Relevant model for $^{40}$K atoms-} To address the nature of ground states for the 4-component mixture of  $^{40}$K atoms we have to break the rotational symmetry associated with the hyperfine spin $f=3/2$. We relabel the states from the $f=9/2$ manifold that were used in experiment \cite{Sengstock} as follows, $m_f=9/2 \to 1/2$, $m_f=1/2\to -1/2$, $m_f=7/2 \to 3/2$, and $m_f=3/2 \to -3/2$, so that the lattice Hamiltonian is again given by Eq.(\ref{model}), but with the modified quartic part, 
\begin{eqnarray}
\label{interactions}
H_{int,j}&=&  \sum_{\alpha<\beta}U_{\alpha,\beta}n_{\alpha,j}n_{\beta,j}  \nonumber\\
&+&U_{\gamma} [\psi_{\frac{1}{2},j}^\dag \psi_{-\frac{1}{2},j}^\dag \psi_{\frac{3}{2},j} \psi_{-\frac{3}{2},j}  +h.c.].
\end{eqnarray}
All interaction constants, $U_{\alpha,\beta}$, are different combinations of scattering lengths $a^K_2$, $a^K_4$, $a^K_6$, and $a^K_8$
that we take from the table \cite{Sengstock} and spin-changing interaction amplitude $U_{\gamma} \sim  \frac{4\pi \hbar^2}{m}(a^K_8-a^K_6)$.
 For $q=0$ and $t\to 0 $ corresponding to infinitely deep lattice, starting from the $t=0$ state and performing degenerate perturbation theory, one can show that a classical (Ising) N{\'e}el state is energetically favorable. The N{\'e}el state breaks translational symmetry with the original $m_f=1/2$ and $m_f=3/2$ components paired on odd sites and $m_f=9/2$ and $m_f=7/2$ paired on even sites (or vice versa). The pairing of $m_f=1/2$ and $m_f=3/2$ components minimizes the interaction energy per lattice site, and due to constraint: $N_{1/2}=N_{9/2}$ and $N_{3/2}=N_{7/2}$ other $L/2$ sites must be occupied by  $m_f=9/2$ and $m_f=7/2$ pairs. Activating $t$ introduces N{\'e}el order due to delocalization energy gain.

\begin{figure}[t]
\vspace*{0.3cm}
\includegraphics[width=8.0cm]{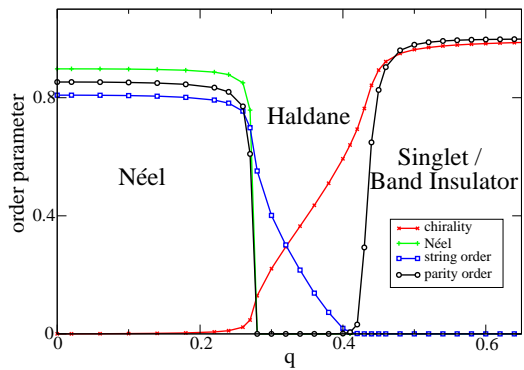}
\caption{(Color online) Quadratic Zeeman coupling (measured in units of $t$) dependence of the order parameters along a cut of Fig.\ref{Fig:Main} for $U/t=20$.}
\vspace*{-0.3cm}
\label{Fig:cut}
\end{figure}

To study effects of quadratic Zeeman coupling we use infinite system size DMRG \cite{Uli} calculations, and monitor different order parameters depicted in Fig.(\ref{Fig:cut}). Explicit expressions of non-local parity and string orders can be borrowed from the corresponding orders of S=1 spin chain \cite{Sebastian} by identifying  N{\'e}el up ( $m_f=1/2$ and $m_f=3/2$ pairs) and down  ($m_f=9/2$ and $m_f=7/2$ pairs) components with  $S^z=+1$ and $S^z=-1$ states, and on-site singlets (initial pairs of BI composed from $m_f=9/2$ and $m_f=1/2$ atoms) with $S^z=0$. The sequences of phase transitions in strong coupling induced by $q$ are shown in Fig.\ref{Fig:cut}. For deep lattices the emergence of  N{\'e}el and Haldane phases in the presence of quadratic Zeeman coupling is a generic feature of 4-component fermions with spin-changing processess allowed for $f\neq 3/2$. Haldane phase can intuitively be understood as a N{\'e}el order diluted with the defects that represent on-site singlets of the BI state. The defects are created on even number of adjacent sites due to the quantum fluctuations on top of the underlying N{\'e}el state.

 For shallow lattices, as depicted in Fig.\ref{Fig:Main}, a dimer state wins. This is consistent with weak-coupling bosonization analysis.
For the scattering lengths taken from the experiment in \cite{Sengstock}, the transition from dimer to N{\'e}el state happens in the strong coupling regime with increasing lattice depth, where bosonization cannot be trusted. However one can induce similar phase transition in weak-coupling by changing interaction coefficients in Eq.(\ref{interactions}). This allows to understand dimer to N{\'e}el transition qualitatively as a Gaussian transition in the $\phi_{t_2}$ sector. In dimer phase the fields are pinned as follows: $\langle \phi_v \rangle = \langle \phi_{t1} \rangle= \langle \phi_{t2} \rangle=  0$, whereas in  N{\'e}el phase:  $\langle \phi_v \rangle = \langle \phi_{t1} \rangle=  0$ and $\langle \phi_{t2} \rangle= \sqrt{\pi}/2$ \cite{Wu06}.
With increasing $q$ an Ising phase transition happens from N{\'e}el to Haldane phase in the $\phi_v$ sector also captured by the effective model of Eq.(\ref{effective}). The expectation values of bosonic fields in Haldane phase are: $\langle \theta_v \rangle = \langle \phi_{t1} \rangle=  0$ and  $\langle \phi_{t2} \rangle= \sqrt{\pi}/2$, and with further increasing $q$ the singlet/BI state should be recovered, with:  $\langle \theta_v \rangle = \langle \phi_{t1} \rangle= \langle \phi_{t2} \rangle=  0$. Hence, occurence of Haldane phase between N{\'e}el and BI is natural from bosonization.
Between Haldane phase and BI (like between dimer and N{\'e}el) a Gaussian criticality from gapless $\phi_{t_2}$ sector is expected, characterized by non-universal exponents. This is in agreement with DMRG results failing to capture universal exponents at these phase transitions and confirming their smooth nature (second order). The complete numerical ground state diagram is presented in Fig.\ref{Fig:Main}.

For $N$-component fermions at half-filling, with attractive interactions and $SU(2)$ spin rotation symmetry, the emergence of charge Haldane insulating phases was shown in \cite{Nonne} and spin Haldane phase was discussed for alkaline-earth atoms in \cite{Totsuka}.

{\it In summary-} a rich ground state phase diagram of half-filled 4-component repulsively interacting alkali fermions can be explored by changing the lattice depth and the quadratic Zeeman coupling, without the need to change the scattering lengths. For the case of $f=3/2$ fermions dimer and singlet phases are realized, whereas when 4 components are from $f>3/2$ multiplet (like $^{40}$K atoms), four distinct gapped MI phases are expected. In particular a topological Haldane phase can be stabilized starting from the BI initial state (with vanishing entropy per particle) and allowing spin changing collisions by adiabatically changing the magnetic field. We hope that our work will motivate experiments on spinor quantum lattice gases to observe spin order beyond the super-exchange scale. These MI phases may be created and detected within the current state-of-the-art techniques in optical lattices. Due to a shallow harmonic trap
along the lattice the MI phases will occupy the central region as in the case of BI \cite{Sengstock}. The different phases may be detected
by different means including monitoring the chirality in time of flight \cite{Sengstock}
and using Faraday rotation as proposed in \cite{Eckert}. 
The string-order of the Haldane phase may be studied using
similar site-resolved measurements such as those recently reported for the measurement of nonlocal parity order in MI \cite{BlochString}.

This work has been supported by QUEST (Center for Quantum Engineering and Space-Time Research) and DFG Research Training Group (Graduiertenkolleg) 1729.

\end{document}